\documentclass{zermatt}
\usepackage{graphicx}
\usepackage{natbib}
\usepackage[T1]{fontenc}
\usepackage[utf8]{inputenc}
\usepackage[english]{babel}
\usepackage{wrapfig}
\def\aap{A\&A}%
\def\apjl{ApJ}%
   \newcommand{\mic}{$\mu$m}

  \newcommand{\smallsub}[1]{{\mbox{{\tiny #1}}}}

  \newcommand{\Av}{A$_\smallsub{V}$}

  \newcommand{\mhmol}{M$_\mathrm{\tiny H_{2}}$}
  
   \newcommand{\hmol}{H$_2$}
  \newcommand{\hi}{H$\,${\sc i}}
  \newcommand{\halpha}{H$\,${$\alpha$}}

  \newcommand{\cplus}{C$^+$}
   
  \newcommand{\cii}{[C$\,${\sc ii}]}
  
  \newcommand{\oiii}{[O$\,${\sc iii}]}

  \newcommand{\ciiline}{\cii$\lambda 158$\mic}

  \newcommand{\oiiilineup}{\oiii$\lambda 88$\mic}

\newcommand{\ciico}{L$_\mathrm{[C\,{\sc II}]}$/L$_\mathrm{CO(1-0)}$}

\newcommand{\ciifir}{L$_\mathrm{[C\,{\sc II}]}$/L$_\mathrm{FIR}$}

\newcommand{\cofir}{L$_\mathrm{[C\,{\sc O}]}$/L$_\mathrm{FIR}$}

\newcommand{\lco}{L$_\mathrm{CO}$}
\newcommand{\lfir}{L$_\mathrm{FIR}$}

\begin{document}
\title{LMC$^+$ SOFIA Legacy Program}
\runningtitle{LMC$^+$ SOFIA Legacy Program}
\author{Suzanne C. Madden and The LMC$^+$ Consortium}
\address{AIM, CEA, Saclay, Gif-sur-Yvette, France 91191,\\ \email{suzanne.madden@cea.fr}}
\begin{abstract}
With the goal of elucidating the effects of low metallicity on the star formation activity, feedback and interstellar medium of low metallicity environments, SOFIA has observed a 40' x 20' (60 pc x 30 pc)  area of our neighboring metal-poor Large Magellanic Cloud in \cii\ and \oiii, targeting the southern molecular ridge. We find extensive \cii\  emission over the region, which encompasses a wide variety of local physical conditions, from bright compact star forming regions to lower density environments beyond, much of which does not correspond to CO structures. Preliminary analyses indicates that most of the molecular hydrogen is in a CO-dark gas component. 

\end{abstract}%
\maketitle
\section{Introduction}
Investigation of the interstellar medium (ISM) conditions for star formation in the very early universe, when metal enrichment was low, is currently raising new and exciting questions and challenging theory, especially in light of the unprecedented results coming out of the new James Webb Space Telescope (JWST) observations showing high excitation emission line properties and moderately metal poor ISM \citep[e.g.][]{schaerer22,curti23}. Some galaxies nearing the Epoch of Reionization seem to show some observational similarities to the local low metallicity (Z), star-forming dwarf galaxies \citep[e.g.][]{cormier19}. With local universe, low-Z laboratories, the closest one being the Large Magellanic Cloud (LMC; d=50kpc), we can directly address questions related to the distribution of the neutral dense, diffuse and ionised gas phases, and properties of ionising sources, while witnessing the propagation of UV photons when the metal abundance is reduced. To study the details of the ISM phases and the feedback from star formation in the highest detail, the SOFIA Joint Legacy Program, LMC$^+$ (co-PIs: S. Madden, A. Krabbe), has mapped the southern molecular ridge in the LMC, just south of 30Doradus, in \ciiline\ and \oiiilineup\ at 2.5pc resolution. Our new SOFIA map is also covered by ALMA in CO (P.I. A. Bolatto), \textit{Herschel} and \textit{Spitzer} continuum bands and many other tracers, allowing us to study the energetics of the ISM well beyond the very limited brightest, star-forming regions which \textit{Herschel} and \textit{Spitzer} spectroscopy have targeted \citep[e.g.][]{lambert_huyghe22}, out into the more extended, diffuse \cplus-emitting gas, where the lower extinction (\Av) conditions can create favorable environments for significant uncharted mass of \hmol.  

\section{Motivation for LMC$^+$}
 Star-forming dwarf galaxies from the Herschel Dwarf Galaxy Surgey \citep[DGS;][]{madden13} have revealed striking observational signatures that set them apart from their metal-rich counterparts. These include: \\
 \hspace{-0.5cm}
 \noindent\underline{\textit{Extreme \ciico\ ratios}}. The consequence of the hard radiation field, together with the low dust abundance, is the widespread ionization and significant photodissociation of CO in the large, weakly-shielded 
\hi-\hmol\ shells of low extinction, resulting in CO more easily photodissociated in dwarf galaxies. We see ratios of \ciico\ in dwarf galaxies often more than a order of magnitude larger than those in star-forming disk galaxies, for example \citep{madden20}. 

\noindent \underline{\textit{Offset from the Kennicutt-Schmidt relationship}}. While star formation surface density correlates with molecular gas surface density for star-forming disk galaxies, star forming dwarf galaxies are systematically offset in the direction suggesting very efficient star formation properties. Molecular gas, traced by CO, is very weak and often not detectable \citep[e.g.][and references within]{cormier15}, eliciting the question: is CO, our usual proxy for molecular gas mass in galaxies, reliably tracing the full mass of \hmol\ (\mhmol) in low-Z galaxies?  

\noindent \underline{\textit{Large reservoirs of CO-dark gas}}. Simulations and multiphase modeling of dwarf galaxies have uncovered significant masses of CO-dark gas \citep[e.g.][and references within]{madden20}.  This phenomenon can exist where \hmol\ is photodissociated by Lyman-Werner band photons which can become optically thick at low \Av\ leading to self-shielded \hmol. The consequence can be a potentially significant reservoir of \hmol\ residing in the \cplus-emitting region, outside of the CO-emitting core.

 \noindent \underline{\textit{Extreme \oiiilineup\ emission}}. With an ionization energy of 35 eV, the \oiiilineup\ is exceptionally luminous in the DGS galaxies, on full galaxy-wide scales \citep{cormier15}. While the \ciiline\ line is normally the brightest FIR fines structure line in star-forming metal-rich galaxies, the \oiiilineup/\ciiline\ ratios in the dwarf galaxies are $>$ 1, and can be as high as 10 \citep{cormier15}, attributed to the significant porosity in dust-poor galaxies \citep{polles19,cormier19,ramambason22}.
 
 \begin{figure}[htp]
\begin{centering}
\vspace{-0.40cm}
\includegraphics[width=0.8\textwidth]{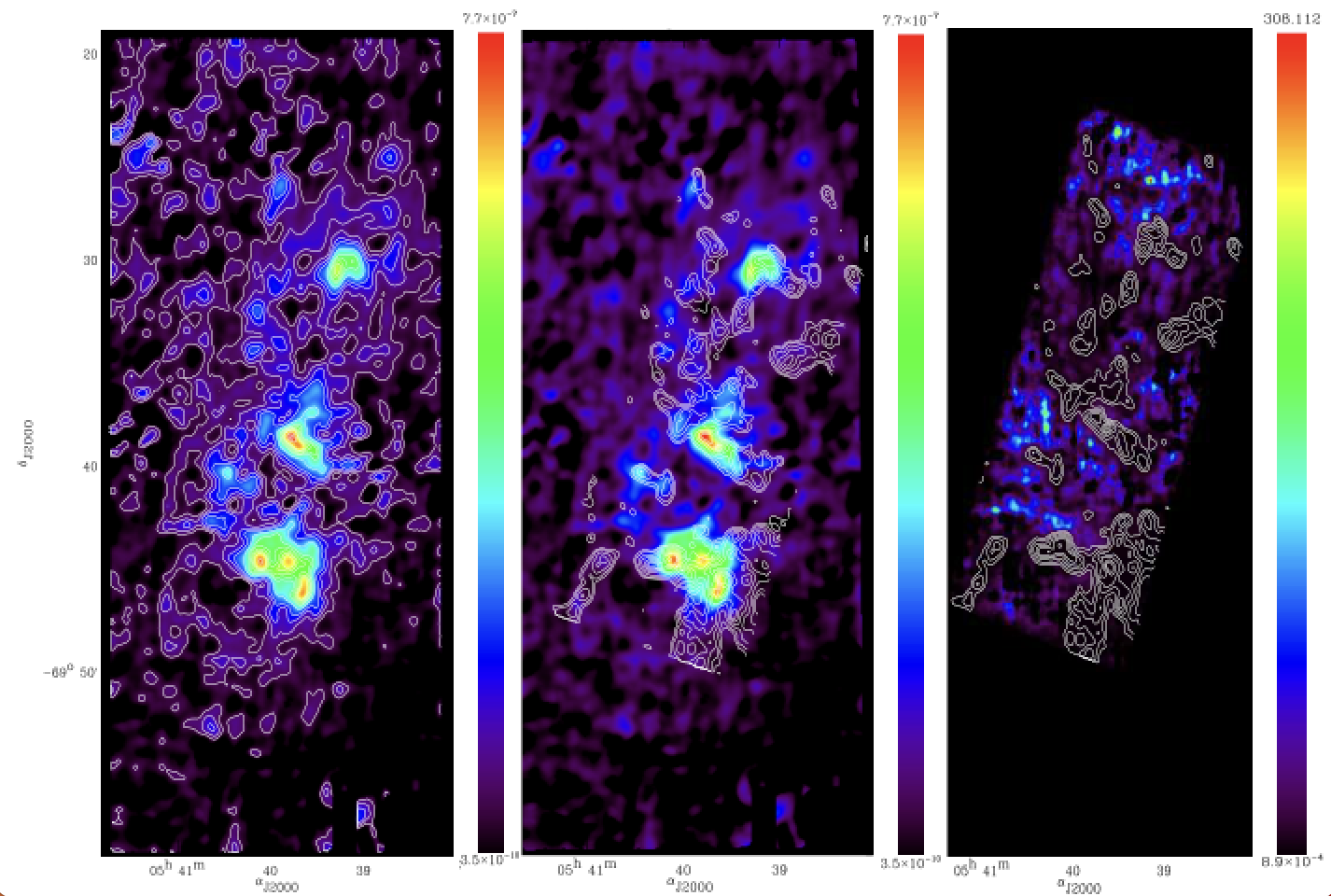}
\caption{SOFIA data from the LMC$^+$ program. (\textit{left}) \cii\ map and contours (color bar: W m$^{-2}$ sr$^{-1}$). Note the star-forming regions, from north to south: N158, N160, N159; (\textit{center}) \cii\ image with CO contours from Tarantino et al. (in prep); (\textit{right}) The image is the ratio \mhmol$_{total}$\ /\mhmol$_{CO}$, which is basically the correction factor from \mhmol\ determined from only CO to total \mhmol\ determined using \cii\ and the conversion from \cite{madden20}. Contours are the CO emission.} 
\label{Madden:fig:co_ciimaps}
\end{centering}

\end{figure}

The main objectives of LMC$^+$ are to determine the physical conditions and thermal processes at work in the photodissociation regions (PDRs), quantify the {\it total} molecular gas mass reservoir throughout this region, and compare the properties of the CO-dark \hmol\ component to those associated with the CO-bright clouds, as well as the atomic and ionized gas distributions.  What local conditions dictate the presence of CO-dark vs. CO-bright reservoirs within galaxies?  In this way we will link an estimate of the total \hmol\ gas mass locally to the star formation rate from YSOs and \halpha, for example. Knowing how to quantify the CO-dark \hmol\ gas mass and comprehend its role in galaxies is absolutely essential in order to correctly determine the full reservoir of molecular gas in galaxies throughout the universe and understand how gas transforms into stars at different cosmological epochs.  We will study the properties of the ISM phases and study how they vary from the proximity of star-forming sites to extended, more diffuse, lower Av phases.

 \section{LMC$^+$ Observations} 
The LMC$^+$ observations were carried out with FIFI-LS onboard SOFIA over 8 flights in march/april 2022 during a deployment to Santiago de Chile. The \cii\ map shows emission peaking toward the compact, bright star-forming complexes, N158, N159 and N160, but also extended emission well beyond and covering much of the map (Figure \ref{Madden:fig:co_ciimaps}).  Comparison of the \cii\ emission with the ALMA CO(2-1) (Tarantino et al. in prep), shows that while peaks of the CO and \cii\ coincide often, many \cii\ structures are not associated with CO emission and vice versa. The map reconstruction of the \oiiilineup\ observations is still ongoing.

  \begin{wrapfigure}{r}{0.5\textwidth}   
 \vspace{-0.2cm}
\label{Madden:fig:ciico}
\hspace{-0.53cm}
    \includegraphics[width=.53\textwidth]{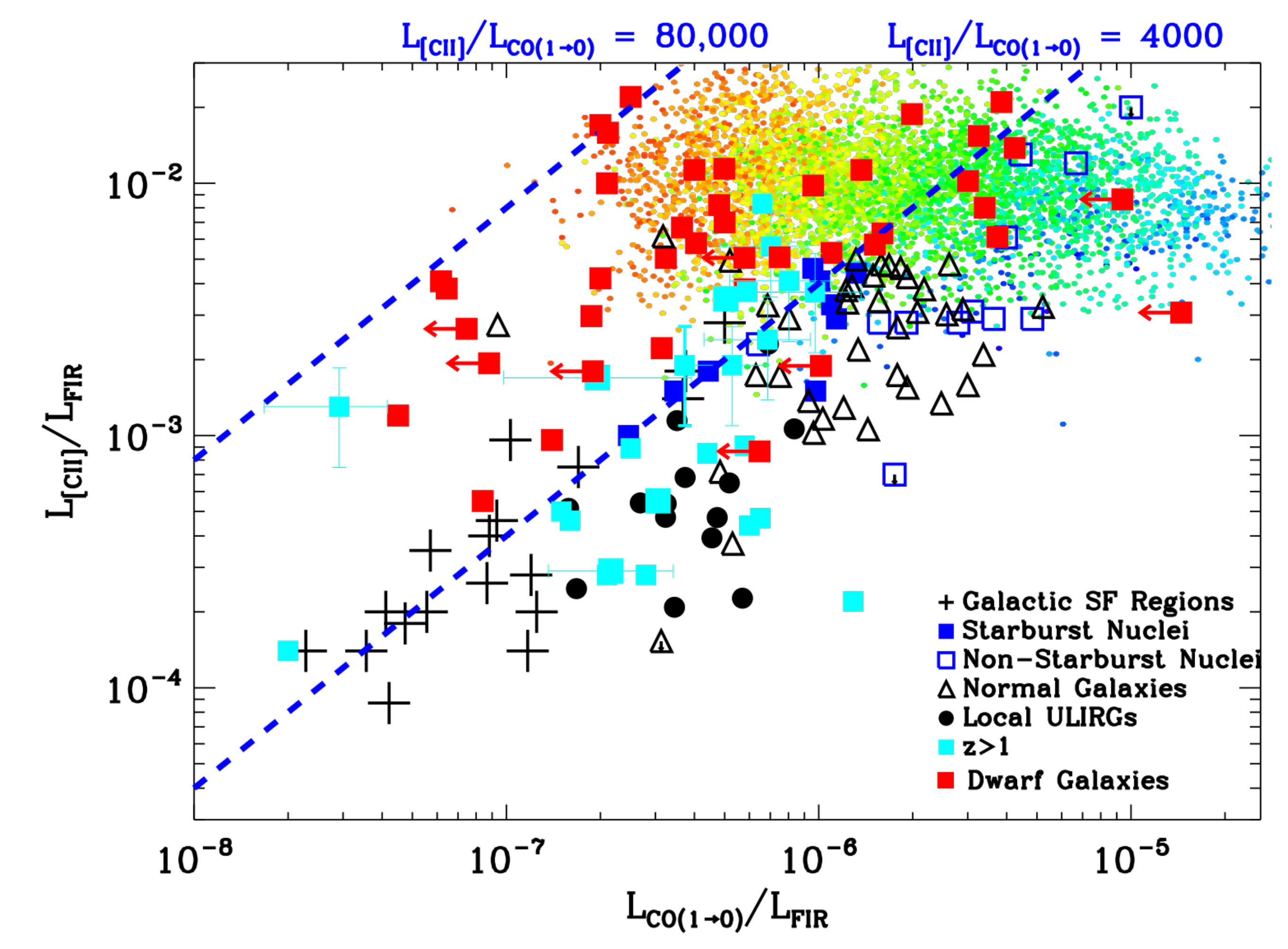}
    \caption{\ciifir\  vs. \cofir\ observed in a variety of galaxies  with dashed lines of constant \ciico\ \citep[see original figure with references in][]{madden20}. Note the location of the low-metallicity dwarf galaxies (red squares) with extreme \ciico\ values. The cloud of small dots are the LMC$^+$ observations with the color sequence ranging from -13.3 (red) to -8.9 (blue) in log CO intensity (W m$^{-2}$ sr$^{-1}$).}
    \vspace{-0.2cm}
    \label{Madden:fig:ciico}
\end{wrapfigure}
 
While the detailed analyses of these maps, along with other complementary data, is ongoing, we attempt a first estimate of the CO-dark gas reservoir and the total \mhmol\ throughout the region.  Our detailed study using Cloudy grids to constrain the multiphase tracers in the Herschel DGS, found that the fraction of CO-dark gas is correlated with \ciico\ resulting in a tight correlation between \cii\ and the total \mhmol\ (Madden et al 2020) and a \cii-to-\mhmol\ conversion factor, which we apply throughout the LMC$^+$ map. Figure \ref{Madden:fig:co_ciimaps} shows the ratio of the \textit{total} \mhmol\ (\mhmol$_{total}$) and \mhmol\ associated with CO only (\mhmol$_{CO}$):  \mhmol$_{total}$\ /\mhmol$_{CO}$, representing the correction factor to convert from the \mhmol\ associated with CO alone to the total \mhmol. These factors can be as high as 100 or more. Regions where CO is emitted (shown in contours) avoid high correction factors, and need little or no correction.
 
The extreme global \ciico\ values observed in low metallicity galaxies suggest the presence of a reservoir of CO-dark gas. Across the LMC$^+$ region, values of \cofir\ range over more than 2 orders of magnitude (Figure \ref{Madden:fig:ciico}), the spread being due, for the most part, to the wide variations of CO, rather than variations in \lfir, as shown by the cloud color sequence. \ciico\ values across the LMC$^+$ map vary over 3 orders of magnitude with high \ciico\ preferring regions with low \lco, holding important implications for CO-dark gas. 

\section{Summary} 
The new SOFIA Legacy program, LMC$^+$, is studying our closest low metallicity galaxy to understand the physical conditions that lead to star formation in molecular clouds, as well as the impact of star formation on the ISM cycle in low metallicity environments. These studies involve multiphase PDR and photoionization modeling and simulations at the scale of giant molecular clouds, to quantify the CO-dark gas reservoir, the geometry and filling factors of the neutral dense, diffuse and ionized gas phases, witnessing the propagation of UV photons when the metal abundance is reduced. LMC$^+$ will contribute to the interpretation of the physical conditions of the ISM and star formation properties in early high-z, low-metallicity galaxies, providing a benchmark paramount for understanding how gas transforms into stars at different cosmological epoch.   

 %
\bibliographystyle{aa}
\bibliography{references}

\end{document}